\documentclass[twocolumn,prl,showpacs,superscriptaddress]{revtex4}
\usepackage{graphicx}
\usepackage{amsmath}
\usepackage{amssymb}
\usepackage{color}

\begin{document}
\title{Fluctuation theorems in general stochastic processes with odd-parity variables}

\author{Hyun Keun Lee}
\affiliation{Department of Physics, University of Seoul, Seoul 130-743, Korea}
\author{Chulan Kwon}
\affiliation{Department of Physics, Myongji University, Yongin, Gyeonggi-Do 449-728, Korea}
\author{Hyunggyu Park}
\affiliation{School of Physics, Korea Institute for Advanced Study, Seoul 130-722, Korea}
\date{\today}

\begin{abstract}
We show that the total entropy production in stochastic processes with odd-parity variables (under time reversal)
is separated into three parts, only two of which satisfy the integral fluctuation theorems in general.
One is the usual excess entropy production, which can appear only transiently and is called nonadiabatic.
Another one is attributed solely to the breakage of detailed balance.
The last part not satisfying the fluctuation theorem comes from the steady-state distribution asymmetry
for odd-parity variables, which is activated in a non-transient manner.
The latter two contributions combine together as the house-keeping (adiabatic) entropy production,
whose positivity is not guaranteed except when the excess entropy production completely vanishes.

\end{abstract}


\maketitle

The (integral) {\it fluctuation theorem} (FT)~\cite{EvansPRL1993,GallavotiPRL1996,CrooksJSP1998,KurchanJPA1998,LebowitzJSP1999}
can be stated for a variable ${\cal R}_r$ (or $\cal R$ in brief) assigned to a random sequence of states (or event) $r$~\cite{EspositoPRL2010} as
\begin{equation}
\label{ft}
\langle e^{-{\cal R}} \rangle \equiv \sum_{r} {\cal P}_r e^{-{\cal R}}=1,
\end{equation}
where ${\cal P}_r$ is the probability of a sequence $r$. As a corollary, the Jensen's inequality guarantees
$\langle {\cal R} \rangle\ge 0$.
Consider $r$ as a path or trajectory in state space, generated during a time interval by a stochastic dynamics.
In case when its {\it functional}  ${\cal R}$~\cite{Onsagar} represents the total entropy production during the process,
the FT has been derived for various nonequilibrium(NEQ) processes, and the thermodynamic 2nd law
$\langle {\Delta S_{\rm tot}} \rangle\ge 0$ automatically follows~\cite{CrooksJSP1998,KurchanJPA1998,JarzynskiPRL1997}.

More recently, Hatano and Sasa found that a part of the total entropy production (excess entropy),
$\Delta S_{\rm ex}$, also satisfies the FT, which represents the entropy production associated with transitions between steady states~\cite{OonoPTP1998,HatanoPRL2001}. Later, Speck and Seifert showed that the remaining part (house-keeping entropy),
$\Delta S_{\rm hk}$, also satisfies the FT, which is required to maintain the NEQ steady state
(NESS)~\cite{SpeckJPA2005,SeifertPRL2005}.
In case of (quasi-static) reversible processes, the system stays at equilibrium almost always during the process, then
the house-keeping entropy production vanishes, $\Delta S_{\rm hk}^{\rm eq}=0$.
Most recently, Esposito {\it et.~al.}~\cite{EspositoPRL2010} interpreted the house-keeping entropy as an adiabatic part
and the excess entropy as a nonadiabatic part of the total entropy production, through a time-scale argument.



Most of findings about the FTs so far hold only
when all state variables have even parity under time reversal, such as position variables.  A typical example is the driven Brownian motion in the over-damped limit. Including odd-parity variables, such as momentum, the mathematical description becomes more complicated in particular for NEQ processes.
Recently, Spinney and Ford suggested a separation of
the total entropy production into three terms for the stochastic system with odd-parity variables~\cite{SpinneyPRL2012}.
The excess entropy production can be cleanly separated out (in fact, exactly the same as
in the case with even-parity variables only) and it satisfies the FT.
However, the house-keeping part composes of two different terms and only one term satisfies the FT.
Especially, the other term not satisfying the FT turns out to be {\em transient}, which
seems inconsistent with the usual adiabatic feature of the house-keeping entropy.
Thus, it was concluded that the physical interpretation of separated entropies is not as clear as in
the even-variable only case (adiabatic vs nonadiabatic), which needs further unraveling.

In this Letter, we present a new scheme of separation for the total entropy production when
odd-parity variables are included. In our scheme, the total entropy production is separated into
the house-keeping and excess contributions, which correspond to the adiabatic and nonadiabatic ones, respectively.
The house-keeping (adiabatic) contribution is composed of two {\em non-transient} characteristic terms,
$\Delta S_{\rm bDB}$ and $\Delta S_{\rm as}$, representing precisely
the detailed balance (DB) breakage and the steady state distribution (SSD) asymmetry for odd-parity variables, respectively.
It is clear that each term is adiabatic with different physical origins. For reversible processes,
it is necessary to require both the DB and the SSD symmetry. Violation of either one brings about
non-vanishing house-keeping entropy production and the processes become irreversible even in the
steady state.
The first term $\Delta S_{\rm bDB}$ obeys the FT, while  does neither the second $\Delta S_{\rm as}$ nor their sum
$\Delta S_{\rm hk}=\Delta S_{\rm bDB}+\Delta S_{\rm as}$.
The nonadiabatic (excess) part $\Delta S_{\rm ex}$ is the same as in the even-variable only case, which satisfies the FT
and so does the total entropy $\Delta S_{\rm tot}=\Delta S_{\rm hk}+\Delta S_{\rm ex}$.


A stochastic process can be described by the master equation
\begin{equation}
\dot p_x(t) = \sum_{y} \omega_{x,y}(\lambda(t))p_y(t),
\label{me}
\end{equation}
where $p_x(t)$ is the probability distribution of state $x$ at time $t$,
and  $\omega_{x,y}$ is the transition rate from $y$ to $x$ for $x\neq y$ with $\omega_{y,y}=-\sum_{x \neq y}\omega_{x,y} (<0)$.
$x$ represents a state vector $(s_1,s_2,\ldots)$ where each component $s_k$ represents a state variable with a definte parity, $\epsilon_k = 1$ (even) or $\epsilon_k = -1$ (odd) under time reversal. The time-reversed state  is
given by $\epsilon x=(\epsilon_1 s_1,\epsilon_2 s_2,\ldots)$.
$\lambda(t)$ denotes a time-dependent protocol as a set of external control parameters.

\begin{figure}
\includegraphics*[width=\columnwidth]{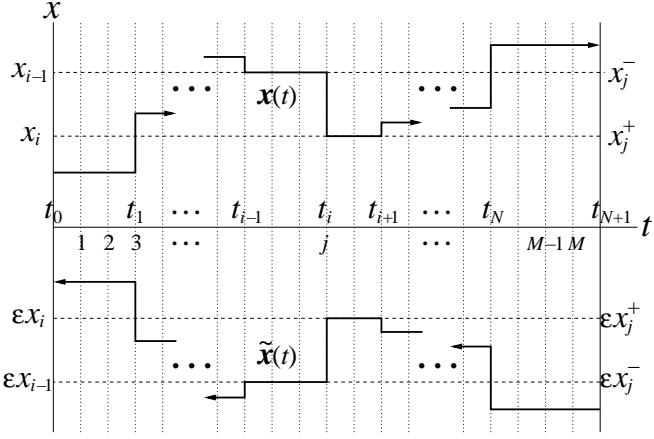}
\caption{
Schematic of a sample path ${\bf x}(t)$ and its time-reversed path $\tilde{\bf x}(\tau-t)$.
The horizontal axis represents time $t$ with $t_0=0$ and
$t_{N+1}=\tau$. The vertical represents state $x$ in the upper and time-reversed state $\epsilon x$ in the lower. There are two time indices. Index $i$ is used for $N$ jumping processes between different states at times $\{t_1,\ldots, t_N\}$. $j$ is used for the time-discretized version such as $t=j\Delta t$ ($j=1,\ldots, M$)
with $\tau=(M+1)\Delta t$ in the $\Delta t\to 0$ limit.  Note that $x_i$ is the state kept unchanged during
a time interval from $t_i$ to $t_{i+1}$.
}
\label{fig1}
\end{figure}

Figure~\ref{fig1} shows a path ${\bf x}(t)$ generated by the master equation with the transition rate matrix
$\omega=\{\omega_{x,y}\}$
during $t=0$ to $\tau$, and
its time-reversed path $\tilde{\bf x}(t)$ defined as ${\bf\epsilon x}(\tau-t)$. We assume that there
are $N$ jumping processes between different states at times $\{t_1,\ldots, t_N\}$.
Then the
probability functional of the ``forward'' path ${\bf x}(t)$
reads
\begin{eqnarray}
\label{opp}
{\cal P}_\omega [{\bf x}]
&\propto&
p_{x_0}\left(
\prod_{i=0}^{N-1} e^{\int_{t_{i}}^{t_{i+1}}dt \omega_{x_{i},x_{i}}(\lambda(t))} \omega_{x_{i+1}, x_{i}}(\lambda_{i+1})
\right)
\nonumber \\
&& \times
e^{\int_{t_{N}}^{\tau}dt \omega_{x_N,x_N}(\lambda(t))}
\end{eqnarray}
where
$p_{x_0}$ is the probability distribution of initial state $x_0$,
$x_i$ is the state for $t_i <t<t_{i+1}$, and $\lambda_i=\lambda(t_i)$.
The time-reversed process is considered under
the protocol changes of $\lambda(t) \rightarrow \lambda(\tau-t)$, and
the initial probability is chosen as the final probability of the forward process, 
$p_{x_N}$.
After a proper rearrangement (see Ref.~\cite{EspositoPRL2010,SpinneyPRL2012} for details), the
probability functional of the ``reverse'' path $\tilde{\bf x}(t)$
reads
\begin{eqnarray}
\label{ropp}
{\cal P}_\omega [\tilde{\bf x}]
&\propto&
p_{x_N}\left(
\prod_{i=0}^{N-1} e^{\int_{t_{i}}^{t_{i+1}}dt \omega_{\epsilon x_{i},\epsilon x_{i}}(\lambda(t))} \omega_{\epsilon x_{i}, \epsilon x_{i+1}}(\lambda_{i+1})
\right)
\nonumber \\
&&~~~~\times e^{\int_{t_N}^{\tau}dt \omega_{\epsilon x_N,\epsilon x_N}(\lambda(t))}~.
\end{eqnarray}
We remark that Eqs.~(\ref{opp}) and (\ref{ropp}) have the same normalization factor since both include the same number of
jumping processes.

The path-dependent total entropy production, $\Delta S_{\rm tot}[{\bf x}]$, is the measure of the irreversibility of
a path ${\bf x}$ with respect to
its time-reversed path $\tilde{\bf x}$, which can be defined as the associated path probability ratio~\cite{SpeckJPA2005,SeifertPRL2005}:
\begin{equation}
\label{stot}
\Delta S_{\rm tot}[{\bf x}] = \ln{{{\cal P}_\omega [{\bf x}]}\over{{\cal P}_\omega [\tilde{\bf x}]}}~.
\end{equation}
Note that $\Delta S_{\rm tot}$ is a FT functional since it satisfies Eq.~(\ref{ft});
$\langle e^{-\Delta S_{\rm tot}} \rangle = \sum_{\bf x} {\cal P}_\omega[{\bf x}] e^{-\Delta S_{\rm tot}}
= \sum_{\tilde{\bf x}} {\cal P}_\omega[\tilde{\bf x}]=1$ (Jacobian $|\partial{\tilde {\bf x}}/\partial {\bf x}|= 1$).
If there are only even-parity variables (all $\epsilon_k=1$), the exponential factors of staying probabilities 
in Eqs.~(\ref{opp}) and (\ref{ropp}) are identical.
These factors are completely canceled out in the probability ratio, and thus only transition rates matter in
$\Delta S_{\rm tot}$. However, it does  not work in that way when odd-parity variables are included, and this is a main source of mathematical difficulty and also different physical origins.

It is convenient to express the path probability
by the conditional probability for transition from $y$ to $x$ during discretized unit time  $\Delta t$ (Fig.~\ref{fig1}),
given as
\begin{equation}
\label{hk}
\Gamma_{x,y}(\lambda(t)) = \delta_{x,y} + \omega_{x,y}(\lambda(t))\Delta t
\end{equation}
where $\delta_{x,y}$ is the Kronecker delta valued $1$ for $x=y$ or $0$ otherwise.
$\Delta t$ is chosen small enough to maintain  $\Gamma_{x,x} > 0$.
Then the two path probabilities can be written as
\begin{equation}
\label{hkpp}
\begin{cases}
{\cal P}_\Gamma[{\bf x}]
= p_{x_0} \prod_{j=1}^M
\Gamma_{x^+_j, x^-_j}(\lambda_j)~,\\
{\cal P}_\Gamma[\tilde{\bf x}]
= p_{x_N} \prod_{j=1}^M
\Gamma_{\epsilon x^-_j, \epsilon x^+_j}(\lambda_j)~,
\end{cases}
\end{equation}
where
$x_j^{+}$ and $x_j^{-}$ represent states just after and before time $t=j\Delta t$ respectively
and $\lambda_j = \lambda(j\Delta t)$.
Note that the product therein includes the staying processes of $x^+_j = x^-_j$
as well as the jumping processes.
Using Eq.~(\ref{hkpp}), one simply writes $\Delta S_{\rm tot}$ as
\begin{equation}
\Delta S_{\rm tot}
=
\Delta S +
\sum_{j=1}^M \ln {{\Gamma_{x^+_j, x^-_j}(\lambda_j)}\over{\Gamma_{\epsilon x^-_j, \epsilon x^+_j}(\lambda_j)}}~,
\label{tot}
\end{equation}
where $\Delta S = -\ln(p_{x_N}/p_{x_0})$ is the entropy change of the system for the forward path. We will later take the $\Delta t \rightarrow 0$ limit to come back to the original problem. The explicit path
dependence of the entropy production is dropped just for simplicity.

The breakage of the DB is an essential characteristics of nonequilibrium processes, which leads to entropy
production even in the NESS. Thus, it would be useful to search for a separation scheme to isolate the entropy production
due to the DB breakage only. The generalized (instantaneous) DB condition at time $t$ for stochastic processes
with odd-parity variables is given as
$\omega_{x,y}(\lambda(t))p_y^s(\lambda(t))=\omega_{\epsilon y, \epsilon x}(\lambda(t))p_{\epsilon x}^s(\lambda(t))$ for $x\neq y$
where $p_x^s(\lambda(t))$ is the SSD of state $x$ for a constant protocol
$\lambda$, whose value
is given by $\lambda(t)$, satisfying the steady state equation
$\sum_x\omega_{y,x}(\lambda)p_x^{\rm s}(\lambda)=0$.
This condition guarantees no physical average currents between states in the steady state
and also yields a relation regarding to the diagonal elements as
$\omega_{x,x}(\lambda(t))p_x^s(\lambda(t))=\omega_{\epsilon x, \epsilon x}(\lambda(t))p_{\epsilon x}^s(\lambda(t))$,
using $\omega_{x,x}=-\sum_{y\neq x} \omega_{y,x}$ .
In terms of the conditional probabilities, the generalized DB condition thus reads as
\begin{eqnarray}
\Gamma_{x,y}(\lambda(t)) &=& \Gamma_{\epsilon y, \epsilon x}(\lambda(t))\frac{p_{\epsilon x}^s(\lambda(t))}
{p_y^s(\lambda(t))}
+\left[1-\frac{p_{\epsilon x}^s(\lambda(t))}{p_x^s(\lambda(t))}\right]\delta_{x,y},\nonumber\\
&=& \delta_{x,y} +\omega_{\epsilon y, \epsilon x}(\lambda(t))\frac{p_{\epsilon x}^s(\lambda(t))}
{p_y^s(\lambda(t))}\Delta t
\label{DB}
\end{eqnarray}

We propose the {\em adjoint} stochastic process with $\Gamma^\dagger_{x,y}$
that can be used to provide a precise measure of the broken DB as
\begin{equation}
\Gamma_{x,y}^\dag(\lambda(t)) = \delta_{x,y} +\omega^\dagger_{x,y}(\lambda(t))\Delta t
\label{gdt}
\end{equation}
with
\begin{equation}
\omega^\dagger_{x,y} =\omega_{\epsilon y, \epsilon x}\frac{p_{\epsilon x}^s}
{p_y^s} .
\label{omegadagger}
\end{equation}
It is trivial to show that $\Gamma^\dag$ is {\it stochastic}
with sufficiently small $\Delta t$~\cite{dt-exp}:
$\sum_x \Gamma^\dag_{x,y} = 1$ and $\Gamma^\dag_{x,y}\ge 0$ for all $x$, $y$.\

When $\Gamma^\dagger_{x,y}=\Gamma_{x,y}$, the DB is satisfied. The entropy production
due to the DB breakage, $\Delta S_{\rm dDB}$, can be defined as
\begin{equation}
\Delta S_{\rm bDB} =
\sum_{j=1}^M \ln \frac{\Gamma_{x^+_j, x^-_j}(\lambda_j)}{\Gamma_{x^+_j, x^-_j}^\dag(\lambda_j)} = \ln{\frac{{\cal P}_\Gamma[{\bf x}]}{{\cal P}_{\Gamma^\dagger}[{\bf x}]}}~,
\label{bDB}
\end{equation}
where ${\cal P}_{\Gamma^\dagger}[{\bf x}]=p_{x_0}\prod_{j=1}^{M}\Gamma_{x^+_j, x^-_j}^\dag(\lambda_j)$ is the
probability of the forward path ${\bf x}$ by the adjoint dynamics. $\Delta S_{\rm bDB}$
is a FT functional by itself, satisfying the integral FT and
must belong to the house-keeping entropy production, since it
contributes even in the steady state.
It also satisfies the detailed FT: $P(R)/P^{\dagger}(-R)=e^{R}$,
where $P(R)$ is the probability that $\Delta S_{\rm bDB} = R$ in the original process while $P^\dag$ is its counterpart in the adjoint process. This is because the mapping to the adjoint dynamics is {\em involutive} ($\Gamma^{\dag\dag}=\Gamma$)~\cite{EspositoPRL2010}, since
both the original and adjoint dynamics share the same SSD ($p^{\rm s}_{x}=p^{\dagger\rm s}_{x}$).




Now subtracting $\Delta S_{\rm bDB}$ from $\Delta S_{\rm tot}$, one can write the remaining part, $\Delta S^\prime=\Delta S_{\rm tot}-
 \Delta S_{\rm bDB}$ as
\begin{equation}
\Delta S^\prime =
\ln\frac{p_{x_0}}{p_{x_N}} + \sum_{j=1}^M \ln \frac{\Gamma_{x^+_j, x^-_j}^\dag(\lambda_j)}{\Gamma_{\epsilon x^-_j, \epsilon x^+_j}(\lambda_j)}~,
\label{mp}
\end{equation}
which is not a FT functional in general, because it is not guaranteed to write down $\Delta S^\prime = \ln {\cal P}_\Gamma[{\bf x}]
/{\cal P}_{\Gamma'}[{\bf x}']$ for the probability functional ${\cal P}_{\Gamma'}[{\bf x}']$ of (reverse) path ${\bf x}'$ in a stochastic dynamics with a certain conditional probability $\Gamma'$. One can find the stochastic condition for $\Gamma^\prime_{y,x} = \Gamma_{x,y}\Gamma_{\epsilon y, \epsilon x}/\Gamma^\dag_{x,y}$ as
\begin{equation}
\label{aug}
\sum_{y}
\Gamma'_{y,x}
=
1+\left({{p_x^{\rm s}-p_{\epsilon x}^{\rm s}}\over{p_{\epsilon x}^{\rm s}}}\right)
\left({{\Gamma_{x,x}^\dag-\Gamma_{x,x}}\over{\Gamma_{x,x}^\dag}}\right).
\end{equation}
This shows that $\Gamma'$ is in general not stochastic due to $\epsilon$ mismatch (note that $\Gamma^\dag_{x,x}$ also includes $\epsilon$). Exceptions when $p_x^{\rm s} = p_{\epsilon x}^{\rm s}$ or $\Gamma^\dag_{x,x} = \Gamma_{x,x}$ will be revisited later.

We can instead extract the excess entropy part by introducing
another stochastic process with $\Gamma^*_{x,y}$ (exactly the same one
as in the even-variable only case) as
\begin{equation}
\Gamma^*_{x,y}(\lambda(t)) = \delta_{x,y} +\omega^*_{x,y}(\lambda(t))\Delta t
\label{mp23}
\end{equation}
with
\begin{equation}
\omega^*_{x,y} =\omega_{y, x}\frac{p_{x}^s}{p_y^s} .
\label{omegadagger}
\end{equation}
Now, we define the excess entropy production,
$\Delta S_{\rm ex}$, as
\begin{eqnarray}
\Delta S_{\rm ex}
&=&
\ln\frac{p_{x_0}}{p_{x_N}} +
\sum_{j=1}^M \ln \frac{\Gamma_{x^+_j, x^-_j}(\lambda_j)}{\Gamma_{x^-_j, x^+_j}^*(\lambda_j)}
= \ln{\frac{{\cal P}_\Gamma[{\bf x}]}{{\cal P}_{\Gamma^*}[\hat{\bf x}]}}
\nonumber\\
&=&
\Delta S+\sum_{j=1}^{M}\ln\frac{p^{\rm s}_{x^+_j}}{p^{\rm s}_{x^-_j}}~,
\label{mp3}
\end{eqnarray}
where the path for the $\Gamma^*$ process is given by $\hat{\bf x}(t) = {\bf x}(\tau-t)$ (time-reversed
without parity change). Of course, $\Delta S_{\rm ex}$ is again a FT functional, satisfying
the integral FT.

The remaining part, $\Delta S_{\rm as}=\Delta S_{\rm tot}-\Delta S_{\rm bDB}-\Delta S_{\rm ex}$, can be
written as
\begin{equation}
\Delta S_{\rm as}=\sum_{j=1}^M \ln
\left[\frac{p^{\rm s}_{\epsilon x_j^+}}{p^{\rm s}_{x_j^+}}
+\delta_{x_j^+,x_j^-}\frac{p^{\rm s}_{x_j^+}-p^{\rm s}_{\epsilon x_j^+}}
{p^{\rm s}_{x_j^+} \Gamma_{\epsilon x_j^+, \epsilon x_j^+}}
\right]~.
\label{mp4}
\end{equation}
One can show easily that this part does not satisfy the FT except
vanishing when there is a SSD symmetry as
\begin{equation}
p^s_{\epsilon x}=p^s_x
\label{asym}
\end{equation}
between {\em mirror} (opposite-parity) states.
This {\em asymmetric} entropy production term is present even in the
absence of external driving  $\lambda (t)$ and also in the NESS (clearly not transient),
so must belong to the house-keeping entropy production.
It therefore follows
\begin{equation}
\label{exhk}
\Delta S_{\rm tot} = \Delta S_{\rm ex} + \Delta S_{\rm hk}
\end{equation}
with $\Delta S_{\rm hk} = \Delta S_{\rm bDB} + \Delta S_{\rm as}$, identified as the total house-keeping entropy production,
which does not obey the FT in general.

The house-keeping entropy production should vanish in the reversible (equilibrium) processes, which implies
that the equilibrium (EQ) condition requires not only the DB but also the symmetry
between the SSD of the mirror states, when odd-parity variables are involved.
These two conditions are independent, and our
two house-keeping contributions, $\Delta S_{\rm bDB}$ and $\Delta S_{\rm as}$, measure precisely
the violation of these two EQ conditions, respectively.

It is worthy of noting that $\Delta S_{\rm bDB}$ and $\Delta S_{\rm as}$ steadily contribute to $\Delta S_{\rm tot}$
in the adiabatic process (or even at $\dot \lambda = 0$) where the time scale of $\lambda(t)$-change is much larger than the relaxation time.
This time scale argument is the reasoning behind the classification of adiabatic and non-adiabatic contributions in $\Delta S_{\rm tot}$, proposed in Ref.~\cite{EspositoPRL2010}. In this criterion, both $\Delta S_{\rm bDB}$ and $\Delta S_{\rm as}$ are the adiabatic contributions while $\Delta S_{\rm ex}$ is the non-adiabatic one. So the total house-keeping
entropy production comprising of $\Delta S_{\rm bDB}$ and $\Delta S_{\rm as}$ is the only and full adiabatic contribution to the total entropy production.


In the $\Delta t \rightarrow 0$ (i.e., $M\rightarrow \infty$) limit, one can obtain
\begin{eqnarray}
\label{r1}
\Delta S_{\rm ex} &=&
\ln\frac{p_{x_0}}{p_{x_N}} +
\sum_{i=1}^N
\ln {{p_{x_{i}}^{\rm s}(\lambda(t_i))}\over{p_{x_{i-1}}^{\rm s}(\lambda(t_i))}}~,
\\
\label{r2}
\Delta S_{\rm hk}&=& \sum_{i=0}^{N} \int_{t_{i}}^{t_{i+1}} dt \left(\omega_{x_{i},x_{i}}(\lambda(t)) - \omega_{\epsilon x_{i},\epsilon x_{i}}(\lambda(t))\right) \nonumber\\
&&+\sum_{i=1}^N \ln {{\omega_{x_i,x_{i-1}}(\lambda(t_i))p^s_{x_{i-1}}}
\over{\omega_{\epsilon x_i, \epsilon x_{i-1}}(\lambda(t_i))p^s_{x_{i}}}}~
\\
\Delta S_{\rm bDB} &=& \sum_{i=0}^{N} \int_{t_{i}}^{t_{i+1}} dt \left(\omega_{x_{i},x_{i}}(\lambda(t)) - \omega^\dag_{x_{i},x_{i}}(\lambda(t))\right) \nonumber\\
&&+\sum_{i=1}^N \ln {{\omega_{x_i,x_{i-1}}(\lambda(t_i))}
\over{\omega_{x_i, x_{i-1}}^\dag(\lambda(t_i))}}~,
\\
\label{r3}
\Delta S_{\rm as} &=& \sum_{i=0}^{N} \int_{t_{i}}^{t_{i+1}} dt
~\omega_{\epsilon x_{i},\epsilon x_{i}}(\lambda(t)) \left({{p_{\epsilon x_{i}}^{\rm s}(\lambda(t))}\over{p_{x_{i}}^{\rm s}(\lambda(t))}}-1\right) \nonumber\\
&&+\sum_{i=1}^N \ln{{p_{\epsilon x_i}^{\rm s}(\lambda(t_i))}\over{p_{x_{i}}^{\rm s}}(\lambda(t_i))} ~.
\end{eqnarray}

$\Delta S_{\rm bDB}$ represents the contribution solely responsible for the DB breakage,
which is the total house-keeping entropy in the absence of odd-parity variables.
While a similar contribution was found by Spinney and Ford~\cite{SpinneyPRL2012}, their term contains what is not directly related to the broken DB. In the meantime, $\Delta S_{\rm as}$ is an odd-variable specific term. It characterizes the asymmetry in the SSD for mirror states. Thus, the asymmetric entropy production serves as another important quantity to measure
the irreversibility of nonequilibrium processes that has not been deeply investigated in the literature. Its importance has recently been recognized by Spinney and Ford~\cite{SpinneyPRL2012}, but their term only exists transiently. In our work,
$\Delta S_{\rm as}$ is shown to exist steadily even for $\dot \lambda = 0$.

We briefly mention on the exceptional cases observed in Eq.~(\ref{aug}), where the total entropy production can be divided into two terms, each of which satisfies the FT. Particulary we consider the case in which $\Gamma_{x,x}^\dag=\Gamma_{x,x}$ (the other case of $p^{\rm s}_{\epsilon x} = p^{\rm s}_{x}$
leads to the conventional separation by $\Delta S_{\rm as}=0$). The condition gives a new stochastic process $\Gamma'_{y,x}$, distinct from $\Gamma^*_{y,x}$
in Eq.~(\ref{mp23}). Then, one readily finds a new separation as
\begin{equation}
\label{hksp}
\Delta S_{\rm tot}
= \Delta S_{\rm bDB} + \Delta S_{\rm mix}
\end{equation}
where
$\Delta S_{\rm mix} = \Delta S_{\rm ex} + \Delta S_{\rm as}$ also satisfies the FT. In the light of physical origin, $\Delta S_{\rm as}$ belongs to the adiabatic entropy production. From the mathematical point of view, however, it operates with nonadiabatic $\Delta S_{\rm ex}$. Moreover, in the adiabatic limit, we have $\Delta S_{\rm hk}$ only, which can be
cleanly separated into two FT functionals. It will be an interesting study to find
an example of this exception.

Finally, we comment on the intriguing feature of $\Gamma^\dag$ in Eq.~(\ref{gdt}).
This is the generalization of the adjoint process in the even-variable only dynamics~\cite{EspositoPRL2010,HatanoPRL2001,SpeckJPA2005}.
However, $\Gamma_{x,x}^\dag = 1+\omega_{\epsilon x, \epsilon x} (\Delta t)
p_{\epsilon x}^{\rm s}/p_x^{\rm s}$ may become negative
for a finite $\Delta t$ when $p_{\epsilon x}^{\rm s} \neq p_x^{\rm s}$.
This situation may be realized when
a finite $\Delta t$ is used as a model parameter or
in numerical study of continuous-time models. In this case, $\Delta t$ smaller
than that used to introduce the original $\Gamma$ is required to make $\Gamma^\dag$ be a stochastic process. If not fulfilled,
$\Delta S_{\rm hk}$ (or $\Delta S_{\rm a}$) can not be divided into $\Delta S_{\rm bDB}$ and $\Delta S_{\rm as}$. There thus exists a upper bound of $\Delta t$ to generate such an entropy production that is generic to the system of interest. The related implication needs further study.

\begin{acknowledgments}
This work was supported by Mid-career Researcher Program through NRF grant (No.~2010-0026627) funded by the MEST.
\end{acknowledgments}



\begin{thebibliography}{99}
\bibitem{dt-exp} $\Gamma^\dag_{x,x}\ge 0$ is not guaranteed unless time is continuous. We will briefly visit this issue in the end.
\end{thebibliography}
\end{document}